\documentclass[twocolumn,showpacs,preprintnumbers,amsmath,amssymb,pra]{revtex4}
\usepackage{amssymb}
\usepackage{graphicx}% Include figure files
\usepackage{dcolumn}% Align table columns on decimal point
\usepackage{bm}
\usepackage{bbm}
\usepackage{amsmath}
\usepackage{subfigure}
\usepackage{epsfig}
\begin{document}
\title{Strongly-coupled Josephson junction array for simulation of frustrated one-dimensional spin models}
\author{Liang-Hui Du}
\author{Xingxiang Zhou}
\email{xizhou@ustc.edu.cn}
\author{Yong-Jian Han}
\author{Guang-Can Guo}
\author{Zheng-Wei Zhou}
\email{zwzhou@ustc.edu.cn}

\affiliation{Key Laboratory of Quantum Information and Department
of Optics and Optical Engineering, University of Science and
Technology of China, Chinese Academy of Sciences, Hefei, Anhui
230026, China }
\date{\today}

\begin{abstract}
We study the capacitance-coupled Josephson junction array beyond the
small-coupling limit. We find that, when the scale of the system is
large, its Hamiltonian can be obtained without the small-coupling
approximation and the system can be used to simulate strongly
frustrated one-dimensional Ising spin problems. To engineer the
system Hamiltonian for an ideal theoretical model, we apply a
dynamical decoupling technique to eliminate undesirable couplings in
the system. Using a 6-site junction array as an example, we
numerically evaluate the system to show that it exhibits important
characteristics of the frustrated spin model.
\end{abstract}

\pacs{03.67.Ac, 75.10.Jm, 85.25.Cp}

\maketitle

\section{Introduction}
As an important application of quantum information science, quantum
simulation of difficult physics problems has received much attention
in recent years. Theoretically, there have been many proposals of
quantum simulators based on various physical systems
\cite{ref:TheoQuanSim1,ref:TheoQuanSim2,ref:TheoQuanSim3,ref:TheoQuanSim4,ref:TheoQuanSim5}.
Experimentally, simulations of some important physics models have
been demonstrated
\cite{ref:QuSimOptLatt1,ref:QuSimOptLatt2,ref:QuSimOptLattMag1,ref:QuSimOptLattMag2,ref:QuSimTrapIon1,ref:QuSimTrapIon2}.

Quantum simulation is most valuable for studying strongly correlated
problems since there are no generally-applicable theoretical methods
to solve them. The strong interactions involved in these problems
usually translate into strong couplings between entities in a
quantum simulator. This requirement of strong couplings often poses
a challenge for the design and construction of a quantum simulator,
since it can be difficult to engineer such couplings in a simulation
system. Even when strong couplings are available, it can still be
nontrivial to tailor the couplings in an well-controlled manner that
is required for the problems to be simulated. As one such example,
it is usually difficult to obtain the system Hamiltonian for a
Josephson junction array when the couplings between the junctions
are strong.  Consequently, Josephson device based quantum simulation
systems often operate in the small coupling limit in which the
system Hamiltonian can be obtained by treating the coupling as
perturbation
\cite{ref:smlCapApprox1,ref:smlCapApprox2,ref:smlCapApprox3}.

In order to go beyond the small-coupling limit and construct a
system useful for simulating strongly correlated physics, in this
paper we investigate a one-dimensional Josephson junction array
which is coupled by large capacitances that cannot be treated
perturbatively. Interestingly, the system Hamiltonian can be
obtained exactly without the small coupling approximation. It is
found that, in the large coupling limit, the interaction strength
between next nearest neighbors can become comparable with that
between the nearest neighbors. Because of this, we can use the
system to study the important problem of one-dimensional frustrated
spin models whose phase diagrams and properties have not been
completely resolved
\cite{ref:FrusModPhaDiaAgr1,ref:PhsDiaInteF,ref:PhsDiaFinitSiz,ref:FrusModPhaDiaAgr2,ref:FrusModPhaDiaAgr3}.
In order to control the system Hamiltonian to match that of the
ideal theoretical model, we use a dynamical decoupling technique to
suppress interactions between neighbors that are three site
locations or farther apart.

\section{System Hamiltonian of the Josephson-junction array}
\begin{figure}[b]
{\includegraphics[width=0.8\columnwidth]{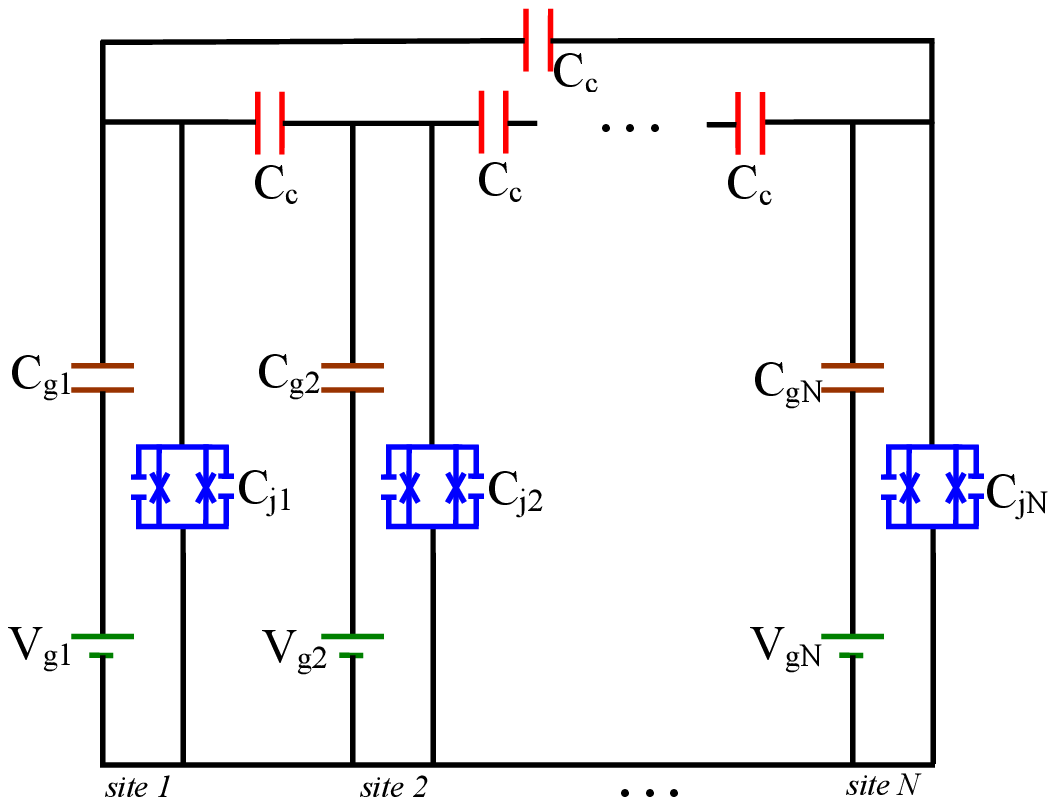}}
\caption {(Color online) The capacitance-coupled Josephson junction
array. $C_c $ is the coupling capacitance between neighboring
junctions. $V_{gi}$ is the bias voltage and $C_{gi}$ is the bias
capacitance of the Josephson junction at site $i$. The total
capacitance and Josephson energy are $C_{ji}$ and $\mathcal
{E}_J$.\label{JoseJunArr}}
\end{figure}

The system we study is an $N$-site Josephson junction array as shown
in Fig. \ref{JoseJunArr}. Each site consists of a charge island
biased by a voltage source $V_{g_i}$ through a gate capacitance
$C_{g_i}$, where $i=1,...,N$. The charge on the island can tunnel
through a SQUID device whose total capacitance is $C_J$ and whose
effective Josephson energy $\mathcal {E}_J$ can be adjusted by a
flux bias. Adjacent charge islands, as well as those at the ends of
the array, are coupled by a capacitance $C_c$. We take the average
phase $\varphi_i$ of the SQUID on site $i$ as the generalized
coordinates for the system. Its rate of change is determined by the
voltage $V_j$ across the Josephson junction according to the
Josephson relation $V_j=(\hbar/2e)\dot{\varphi}_i$
\cite{ref:SupConduct}. The charging energy $T$ of the capacitances
and the Josephson energy $V$ of the junctions are
\begin{eqnarray}
T&=&\frac{1}{2}\sum\limits_{i=1}^{N}[(\frac{\hbar}{2e})^2C_{J}\dot{\varphi}_i^2+C_{gi}(V_{gi}-\frac{\hbar}{2e}\dot{\varphi}_i)^2\nonumber\\
&+&C_c(\frac{\hbar}{2e})^2(\dot{\varphi}_i-\dot{\varphi}_{i+1})^2].\label{kiEn}\\
V&=-&\sum_i\mathcal {E}_J\cos\varphi_i.\label{poEn}
\end{eqnarray}

From the Lagrangian $\mathcal {L}=T-V$ of the system, we can
derive the generalized momentum which is related to the charge
number on the islands
\begin{equation}
P_i=\partial\mathcal {L}/\partial\dot{\varphi_i} \equiv -\hbar
n_i.\label{generalmomentum}
\end{equation}

Denoting the charge number on the $i$th island $n_i$ and the bias
charge number $n_{gi}=C_{g_i}V_{g_i}/2e$, we can write the system's
equation of motion
\begin{equation}
\frac{\hbar}{(2e)^2}\bm{M}\vec{\dot{\bm{\varphi}}}=\vec{\bm{n}},
\end{equation}
where
$\vec{\dot{\bm{\varphi}}}=[\dot{\varphi}_1,\dot{\varphi}_2,...,
\dot{\varphi}_i,...,\dot{\varphi}_N]^T$,
$\vec{\bm{n}}=[n_1-n_{g1},n_2-n_{g2},...,n_i-n_{gi},...,n_N-n_{gN}]^T$,
and $\bm{M}$ is the matrix
\begin{equation}
\bm{M}=\left(\begin{array}[c]{cccccc}%
C_{\Sigma} & -C_c & \cdots&\cdots&\cdots& -C_c\\
-C_c & C_{\Sigma} & -C_c & \cdots &\cdots& 0\\
0 & -C_c & C_{\Sigma} & -C_c & \cdots & 0\\
\multicolumn{6}{c}{\dotfill}\\
\cdots&\cdots&-C_c&C_{\Sigma }&-C_c&\cdots\\
-C_c&\cdots&\cdots&\cdots&\cdots&\vdots
\end{array}\right),
\end{equation}
where $C_{\Sigma}=C_{J}+C_{g_i}+2C_c$ is the total capacitance of
each island.

The Hamiltonian of this Josephson junction array is:
\begin{eqnarray}
H&=&\sum\limits_{i} P_i\dot{\varphi}_i-\mathcal {L}\nonumber\\
 &=&\frac{1}{2}(\frac{\hbar}{2e})^2\vec{\dot{\bm{\varphi}}}^T\bm{M}\vec{\dot{\bm{\varphi}}}+V\nonumber\\
 &=&\frac{1}{2}(2e)^2\vec{\bm{n}}^T\bm{M}^{-1}\vec{\bm{n}}+V.\label{Hamivecn}
\end{eqnarray}

In order to obtain the system Hamiltonian, the inverse matrix of
$\bm{M}$ must be calculated. Since it is nontrivial to exactly
solve for $\bm{M}^{-1}$, most previous studies
\cite{ref:smlCapApprox1,ref:smlCapApprox2} have assumed a small
coupling capacitance $C_c$ so the system Hamiltonian can be obtained
by treating the coupling as perturbation. Unfortunately, this
precludes the system from being used to simulate strongly-correlated
problems in which strong couplings are required. Here, we assume
that $C_c$ is not necessarily small and try to solve for
$\bm{M}^{-1}$ exactly.

Considering the translational symmetry of the problem, we see the
inverse matrix $\bm{M}^{-1}$ must be in the form
\begin{equation}
\bm{M}^{-1}=\left(\begin{array}[c]{ccccccc}%
a_1   & a_2 & \cdots& a_{N/2+1}&\cdots& a_3 &a_2\\
a_2   & a_1 &    &   &   &  & a_3\\
\vdots & &\ddots &   &   &  & \\
a_{N/2+1}& &   &   &   &      &\vdots\\
\vdots &   &   &   &\ddots &      &\\
a_3    &   &   &   &   &a_1&a_2\\
a_2    &   &   & \cdots&   & a_2  &a_1
\end{array}\right),
\end{equation}

Using some mathematic techniques for solving polynomials, we can
calculate the values of the matrix elements in $\bm{M}^{-1}$
exactly:
\begin{equation}
a_i=\frac{1}{C_{\Sigma}}(\lambda^{i-1}A_0+\frac{1}{\lambda^{i-1}}B_0),\label{elemofInvM}
\end{equation}
where
\begin{eqnarray}
A_0&=&\frac{1}{1-2\beta\lambda+(1-\frac{2\beta}{\lambda})\frac{2\beta-\lambda}{1-2\beta\lambda}\lambda^{N-1}},\label{paraofInvM1}\\
B_0&=&\frac{(2\beta-\lambda)\lambda^{N-1}}{1-2\beta\lambda}A_0,\label{paraofInvM2}\\
\lambda&=&\frac{1-\sqrt{1-4\beta^2}}{2\beta},\label{paraofInvM3}\\
\beta&=&\frac{C_c}{C_{\Sigma}}=\frac{C_c}{2C_c+C_{gi}+C_{J}}<\frac{1}{2}.\label{paraofInvM4}
\end{eqnarray}

\begin{figure}[b]
{\includegraphics[width=0.8\columnwidth]{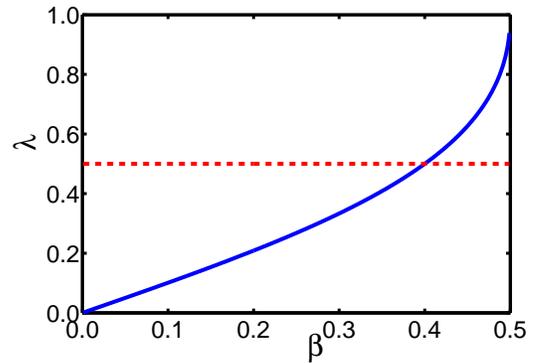}}
\caption {(Color online) The relationship between $\lambda$ and
$\beta$. The dashed red line corresponds to $\lambda=1/2$.
\label{lambdabeta}}
\end{figure}

When $N\rightarrow \infty$, the above results simplify to
$A_0\rightarrow1/(1-2\beta\lambda)$, $B_0\rightarrow0$, and
$a_i=\lambda^{i-1}A_0/C_{\Sigma}$.

As can be seen in Eq. (\ref{Hamivecn}), $\lambda$ characterizes
the ratio between adjacent and non-adjacent interaction strengths
in our system. According to Eqs. (\ref{paraofInvM3}) and
(\ref{paraofInvM4}), $\lambda$ is determined by the coupling
capacitance $C_c$. In the weak coupling limit $C_c \ll
C_{\Sigma}$, $\beta\ll 1$ and $\lambda$ is nearly equal to
$\beta$. However, when the coupling is strong,  $\lambda$
increases quickly, as shown in Fig. \ref{lambdabeta}. In
particular, when the coupling capacitance $C_c$ dominates,
$\lambda$ can approach 1, and the Hamiltonian in Eq.
(\ref{Hamivecn}) describes a deeply frustrated system with
appreciable non-adjacent interactions. Meanwhile, according to the
results of Ref. \cite{ref:LargeCapCoup}, there exist many close
energy levels as $\beta$ approaches 1/2 too closely, which can
fail the two-level approximation. Therefore we only pay our
attention to a large $\beta$, but not closely approaching 1/2 in
the rest of our paper.

Under proper conditions, if we bias the charge islands at the
vicinity of $n_{gi}=1/2$, we can use the two-level approximation for
the charge qubits with $n_i=0$ and $n_i=1$ as the basis states. We
can then write the system Hamiltonian in the following Pauli matrix
representation
\begin{equation}
H_{JJA}=\sum_i\sum_j(-1)^j(\lambda)^{j-1}\sigma_i^z\sigma_{i+j}
^z-B\sum_i\sigma_i^x,
\label{eq:Spin-H}
\end{equation}
where the spin up and down states represent the $n_i=0$ and
$n_i=1$ states, and the transverse magnetic field
$B\!\!=\!\!-\mathcal{E}_jC_{\Sigma}/(2\lambda e^2 A_0)$. $B$ can
be adjusted by tuning the magnetic flux of the SQUIDs. Here we
applied a canonical $\sigma^x$ transformation on even sites for
the convenience of our following discussion about phase diagram
without changing the physics of the system. Usually, two-level
approximation works very well for a single Josephson charge island
\cite{ref:twoLevelAppro1,ref:twoLevelAppro2,ref:twoLevelAppro3}.
In our system of Josephson junction array, more careful analysis
is necessary. Appendix A provides a detailed discussion about the
applicability of the two-level approximation in our system.

\section{Simulation of the ANNNI model using dynamical decoupling}
Our circuit is useful for quantum simulation of frustrated spin
problems since it exhibits strong non-adjacent spin interactions
in the large coupling limit. However, the Hamiltonian in Eq.
(\ref{eq:Spin-H}) does not correspond to an ideal theoretical
model with limited-range interactions yet. We intend to further
engineer it for quantum simulation of well known frustrated
models. As an example, we show how to simulate the one dimensional
axial next-nearest-neighbor Ising (ANNNI) model in external
fields. Its Hamiltonian is given by
\begin{equation}
H_{AI}=-\sum_i\sigma_i^z\sigma_{i+1}^z+\lambda\sum_i\sigma_i^z\sigma_{i+2}^z-B\sum_i\sigma_i^x.
\label{eq:ANNNI-H}
\end{equation}
The ANNNI problem is an important model for studying frustrated
physics due to competition between adjacent and non-adjacent
neighbor interactions \cite{ref:ANNNI1,ref:ANNNI2}. Despite years
of research, its phase diagrams and physical properties have not been
completely resolved
\cite{ref:FrusModPhaDiaAgr1,ref:PhsDiaInteF,ref:PhsDiaFinitSiz,ref:FrusModPhaDiaAgr2,ref:FrusModPhaDiaAgr3}.

\begin{figure}[t]
{\includegraphics[width=0.8\columnwidth]{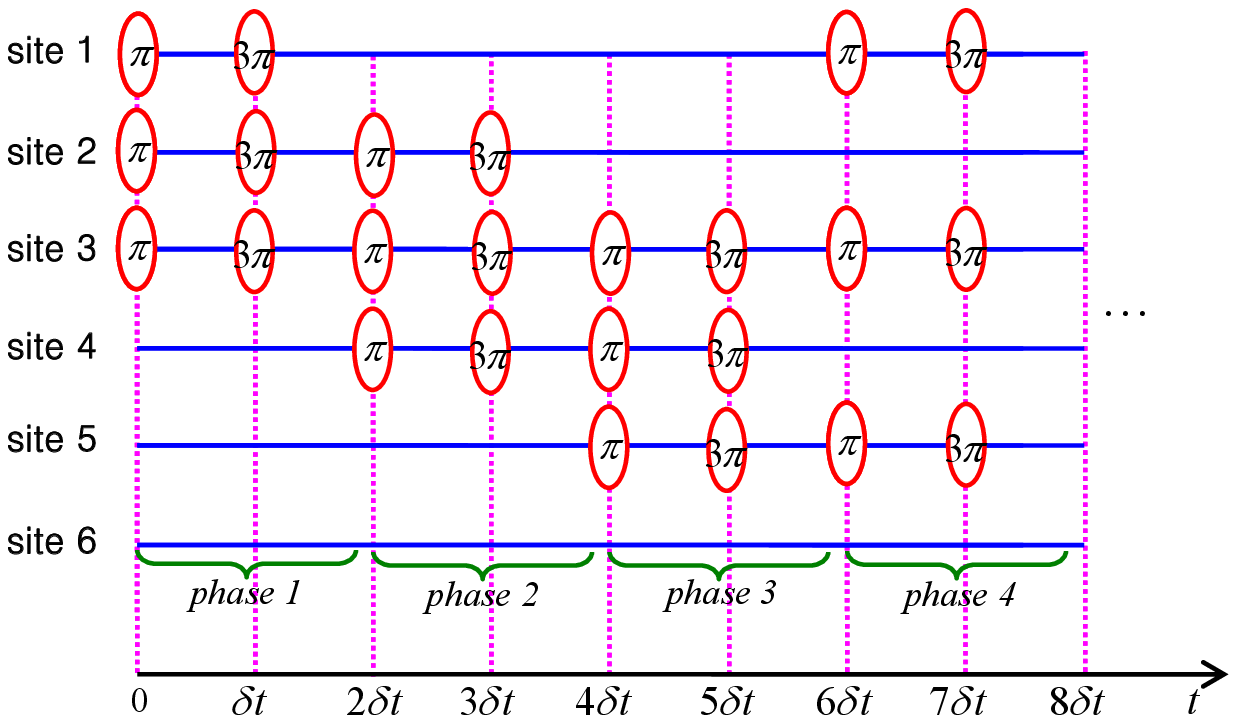}}
\caption{(Color online) The dynamical decoupling control scheme to
eliminate interactions between spins that are 3 sites apart. The
ellipses in groups of 3 along the vertical direction represent
simultaneous $\pi$ or $3\pi$ pulses applied to the qubits.
\label{dd}}
\end{figure}

Comparing our circuit Hamiltonian in Eq. (\ref{eq:Spin-H}) and
target Hamiltonian in Eq. (\ref{eq:ANNNI-H}), we find that there are
extra terms that describe interactions between spins that are 3 or
more sites apart. We eliminate these extra terms by using techniques
of dynamical decoupling \cite{ref:DDtech1,ref:DDtech2}. In this
practice, we apply carefully designed sequences of fast short pulses
to individual qubits to engineer a Hamiltonian that can be very
different than the original Hamiltonian.

In the following, we will demonstrate how to eliminate interactions
between spins that are 3 sites apart for a spin chain with the
Hamiltonian in Eq. (\ref{eq:Spin-H}). For clarification and ease of
illustration, we discuss the details of our scheme on a 6-site spin
chain with periodic boundary condition. The same technique applies
in a spin chain with arbitrary length.

When the number of qubits in the system is 6, the Hamiltonian in Eq.
(\ref{eq:Spin-H}) reads
\begin{eqnarray}
H_{s}&=&J_1(\sigma_1^z\sigma_2^z+\sigma_2^z\sigma_3^z+\sigma_3^z\sigma_4^z+\sigma_4^z\sigma_5^z+\sigma_5^z\sigma_6^z+\sigma_6^z\sigma_1^z)\nonumber\\
     &
&+J_2(\sigma_1^z\sigma_3^z+\sigma_2^z\sigma_4^z+\sigma_3^z\sigma_5^z+\sigma_4^z
\sigma_6^z+\sigma_5^z\sigma_1^z+\sigma_6^z\sigma_2^z)\nonumber\\    &
&+J_3(\sigma_1^z\sigma_4^z+\sigma_2^z\sigma_5^z+\sigma_3^z\sigma_6^z)
-B\sum\limits_{i=1}^6\sigma_i^x
\label{HGenl_six}
\end{eqnarray}
where $J_1$ and $J_2$ are interaction strengths between the nearest
and next-nearest neighbors and $J_3$ is that between spins that are
3 sites apart.

Our scheme to engineer the desired Hamiltonian involves applying rapid pulsed
operations $R_x(\theta )=\exp\{-(i\theta \sigma _x)/2\}$ on individual qubits.
When $\theta=\pi$ ($3\pi$) the corresponding unitary operation is
$R_x(\pi)=-i\sigma^x$
($R_x(3\pi)=R_x^\dagger(\pi)=i\sigma^x$). We have the following
commutation relations:
\begin{eqnarray}
R^\dagger_x(\pi)\sigma^zR_x(\pi)&=&-\sigma^z,\label{commut_relation1}\\
R^\dagger_x(\pi)\sigma^xR_x(\pi)&=&\sigma^x,\label{commut_relation2}\\
R^\dagger_x(\pi)e^{-i\hat{H}t}R_x(\pi)&=&e^{-i[R^\dagger_x(\pi)\hat{H}R_x(\pi)]t}.\label{quan_Alg}
\end{eqnarray}

Our procedure contains four phases as shown in Fig. \ref{dd}. Each
phase is completed in an interval of $2\delta t$ where $\delta t$ is
a short time. At the beginning of each phase, the operations
$R^{(i,j,k)}_x(\pi)\!=\!R^i_x(\pi)R^j_x(\pi)R^k_x(\pi)$
($i,j,k=1,2,...,6$) are applied to the qubits on sites $i,j$, and
$k$ simultaneously (see Fig. \ref{dd}). After a time interval of
$\delta t$, the conjugate operations $R^{\dagger(1,2,3)}_x(\pi)$ are
applied in the second half of the phase. The evolution of the system
at the end of phase 1 is given by
\begin{equation}
U_1=e^{-iH_s\delta t}R^{\dagger(1,2,3)}_x(\pi)e^{-iH_s\delta
t}R^{(1,2,3)}_x(\pi).\label{U_1}
\end{equation}
For a short time duration $\delta t$, we have $e^{-iH_1\delta
t}e^{-iH_2\delta t}=e^{-i(H_1+H_2)\delta t}+O(\delta t^2)$. To first
order in $\delta t$, the unitary operator
$U_1=\exp(-iH_1^{eff}2\delta t)$, where the effective Hamiltonian in
phase 1 is
\begin{eqnarray}
H_1^{eff}&=&H_s+R^{\dagger(1,2,3)}_x(\pi)H_sR^{(1,2,3)}_x(\pi)\nonumber\\
         &=&J_1(\sigma_1^z\sigma_2^z+\sigma_2^z\sigma_3^z+\sigma_4^z\sigma_5^z+\sigma_5^z\sigma_6^z)\nonumber\\
         & &+J_2(\sigma_1^z\sigma_3^z+\sigma_4^z\sigma_6^z)-B\sum\limits_{i=1}^6\sigma_i^x.\label{H1_eff}
\end{eqnarray}

From Eq. (\ref{H1_eff}), we can see that all couplings between
qubits that are 3 sites apart have been eliminated. However, some
terms that we want to keep, such as the nearest-neighbor coupling
$\sigma_3^z\sigma_4^z$ and next-nearest coupling
$\sigma_3^z\sigma_5^z$, are also eliminated. In order to make up for
this problem, in phase 2 and 3 we use the same technique but shift
the target qubits one site a time as shown in Fig. \ref{dd}. This
gives us the following effective Hamiltonian for these two phases
\begin{eqnarray}
H_2^{eff}
&=&J_1(\sigma_2^z\sigma_3^z+\sigma_3^z\sigma_4^z+\sigma_5^z\sigma_6^z+\sigma_6^z
\sigma_1^z)\nonumber\\
&&+J_2(\sigma_2^z\sigma_4^z+\sigma_5^z\sigma_1^z)-B\sum\limits_{i=1}
^6\sigma_i^x;\label{H2_eff}\\
H_3^{eff}
&=&J_1(\sigma_1^z\sigma_2^z+\sigma_3^z\sigma_4^z+\sigma_4^z\sigma_5^z+\sigma_6^z
\sigma_1^z)\nonumber\\
         &
&+J_2(\sigma_3^z\sigma_5^z+\sigma_6^z\sigma_2^z)
-B\sum\limits_{i=1}^6\sigma_i^x.\label{H3_eff}
\end{eqnarray}
Obviously, the missing terms for nearest and next-nearest neighbor
couplings in $H_1^{eff}$ in Eq. (\ref{H1_eff}) are compensated by
remaining terms in Eqs. (\ref{H2_eff}) and (\ref{H3_eff}).
Similarly, missing terms in $H_2^{eff}$ and $H_3^{eff}$ are
compensated. Nevertheless, some next-nearest neighbor coupling terms
are still missing from the sum of $H_1^{eff}$, $H_2^{eff}$ and
$H_3^{eff}$, since in each phase 2/3 of the nearest-neighbor
couplings remain but only 1/3 of the next-nearest-neighbor couplings
survive. In order to obtain all the nearest-neighbor and the
next-nearest-neighbor couplings, we need a phase 4 as shown in Fig.
\ref{dd}. By keeping the next-nearest-neighbor interactions and
eliminating the nearest-neighbor interactions, it gives us the
effective Hamiltonian
\begin{eqnarray}
H_4^{eff}&=&J_2(\sigma_1^z\sigma_3^z+\sigma_2^z\sigma_4^z+\sigma_3^z\sigma_5^z+\sigma_4^z\sigma_6^z+\sigma_5^z\sigma_1^z+\sigma_6^z\sigma_2^z))\nonumber\\
         & &-B\sum\limits_{i=1}^6\sigma_i^x.\label{H4_eff}
\end{eqnarray}

The combined evolution of the system for the 4 phases is
\begin{eqnarray}
U\!\!&=&\!\!U_4U_3U_2U_1\!\approx \!e^{-iH_4^{eff}2\delta
t}e^{-iH_3^{eff}2\delta
t}e^{-iH_2^{eff}2\delta t}e^{-iH_1^{eff}2\delta t}\nonumber\\
&\approx&\!\!e^{-i[H_4^{eff}+H_3^{eff}+H_2^{eff}+H_1^{eff}]2\delta
t}\nonumber\\
&\!\!\triangleq&e^{-iH_{eff}8\delta t}
\end{eqnarray}
where the effective average Hamiltonian is
\begin{eqnarray}
H_{eff}&=&\frac{1}{2}[J_1(\sigma_1^z\sigma_2^z+\sigma_2^z\sigma_3^z+\sigma_3^z\sigma_4^z+\sigma_4^z\sigma_5^z+\sigma_5^z\sigma_6^z+\sigma_6^z\sigma_1^z)\nonumber\\
      & &+J_2(\sigma_1^z\sigma_3^z+\sigma_2^z\sigma_4^z+\sigma_3^z\sigma_5^z+\sigma_4^z\sigma_6^z+\sigma_5^z\sigma_1^z+\sigma_6^z\sigma_2^z)]\nonumber\\
      & &-B\sum\limits_{i=1}^6\sigma_i^x.\label{H_eff}
\end{eqnarray}
This is exactly the ANNNI model with periodic boundary
condition.

Though we have used a 6-site chain to demonstrate how to eliminate
couplings between qubits that are 3 sites apart, it is obvious that,
by applying the operations $R^{(i,j,k)}_x(\pi)$ to groups of 3
qubits in the chain and shifting the target qubits by 1 site at a
time in each phase, we can use the same technique to eliminate
couplings between qubits 3 sites apart in an infinite-length spin
chain. Notice that in Eq. (\ref{eq:Spin-H}) there are couplings
between qubits 4 or more sites apart. These couplings are weaker
since the interaction strengths $\lambda^{j-1}$ in Eq.
(\ref{eq:Spin-H}) decreases with the distance between qubits, but
the error caused by them may still be unacceptable depending on the
required accuracy of the simulation. By using a nested dynamical
decoupling scheme, it can be shown that all couplings between qubits
that are separated by 3 or more sites can be eliminated \cite{Du}.
Therefore, given a required accuracy, we can in principle achieve
the ANNNI Hamiltonian in Eq. (\ref{eq:ANNNI-H}).

\section{phase diagram of the ANNNI model}
Now that we can simulate the ANNNI model, we perform some
analysis on its phase diagram. When $B=0$ and $\lambda$ is small,
the nearest-neighbor interactions dominate and the ground state is
the ferromagnetic state
$|\downarrow\downarrow\cdots\downarrow\downarrow\cdots\rangle_z$
(or $|\uparrow\uparrow\cdots\uparrow\uparrow\cdots\rangle_z$). As
$\lambda$ increases, the next-nearest-neighbor interactions become
important. When $\lambda$ reaches some critical value, they become
the dominating factor and the antiphase
$|\uparrow\uparrow\downarrow\downarrow\uparrow\uparrow\cdots\rangle_z$
(or
$|\downarrow\downarrow\uparrow\uparrow\downarrow\downarrow\cdots\rangle_z$)
which minimizes the next-nearest neighbor interaction energy
becomes the ground state. In the limit of large transverse field
$B$, the ground state will be the paramagnetic phase
$|\uparrow\uparrow\cdots\uparrow\uparrow\cdots\rangle_x$ to
minimize the Zeeman energies.

From the above analysis, we see that there should be a ferromagnetic phase, a
paramagnetic phase and an antiphase in the ANNNI model. However, there could be
more subtle regimes in the phase diagram. Studies have shown (inconclusively)
that there could be a unique floating phase in the deeply frustrated regime
\cite{ref:FrusModPhaDiaAgr2,ref:FrusModPhaDiaAgr3}. This phase is
characterized by the fact that the $n$th-neighbor spin-spin correlation function
in the longitudinal direction decays algebraically. The exact origin and range
of the floating phase is still an
open question and therefore a good subject for quantum simulation. Since the
floating phase is located in the deeply frustrated regime, our circuit with
strong couplings offers a good system for its simulation.

Many numerical recipes such as finite-size scaling method
\cite{ref:PhsDiaFinitSiz} and the interface approach
\cite{ref:PhsDiaInteF} have been used to calculate the phase
diagram of frustrated Ising model. We use the new method of
time-evolving block decimation (TEBD) algorithm
\cite{ref:TEBDMeth1,ref:TEBDMeth2} to calculate the ground state
energy of ANNNI model. TEBD is an powerful algorithm to simulate
quantum evolution process based on matrix product state and
Trotter expansion \cite{ref:TEBDMeth1,ref:TEBDMeth2}. By making
the evolution time imaginary, we get the so called i-TEBD which
can be used to determine the ground state of a system efficiently.
The TEBD method is based on the following matrix representation of
a quantum state
\begin{equation}
|\Psi\rangle=\sum_{i_1=1}^d\cdots\sum_{i_n=1}^dc_{i_1\cdots
i_n}|i_1\rangle\otimes\cdots\otimes|i_n\rangle
\label{MPS}
\end{equation}
where $d$ is the number of
local energy levels on every site. The coefficients
\begin{equation}
c_{i_1\cdots
i_n}\!=\!\!\!\sum_{\alpha_1,\cdots,\alpha_{n-1}}^\chi\!\!\!
\Gamma_{\alpha_1}^{[1]i_1}\xi
_{\alpha_1}^{[1]}\Gamma_{\alpha_1\alpha_2}^{[2]i_2}
\xi_{\alpha_2}^{[2]}\Gamma_{\alpha_2\alpha_3}^{[3]i_3}\cdots\Gamma_{\alpha_{n-1}}^{[n]i_n}.\label{coefMPS}
\end{equation}
are defined with the help of $n$ tensors
$\{\Gamma^{[1]},\cdots,\Gamma^{[n]}\}$ and $n-1$ vectors
$\{\xi^{[1]},\cdots,\xi^{[n-1]}\}$, where $\chi$ is the maximal
number of two-party Schmidt decomposition coefficients. In practice,
$\chi$ does not need to be very large, because the Schmidt
coefficients roughly decay exponentially with $\alpha$. Any
single-site operation or adjacent-site joined operation on the state
can be achieved by updating the corresponding tensors and vectors.
Here, we use the second order Trotter expansion for i-TEBD (see
reference \cite{ref:TEBDMeth2}).

\begin{figure}[h]
{\includegraphics[width=0.8\columnwidth]{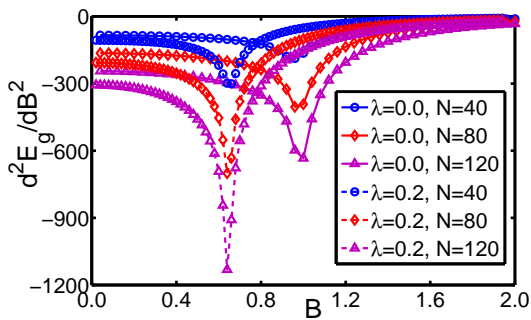}} \caption
{(Color online) The finite-size scaling of second order derivative
of ground state energy with $\lambda\doteq0$ (solid lines) and
$\lambda=0.2$ (dashed lines) \label{ProPT}}
\end{figure}

\begin{figure}[t]
{\includegraphics[width=0.7\columnwidth]{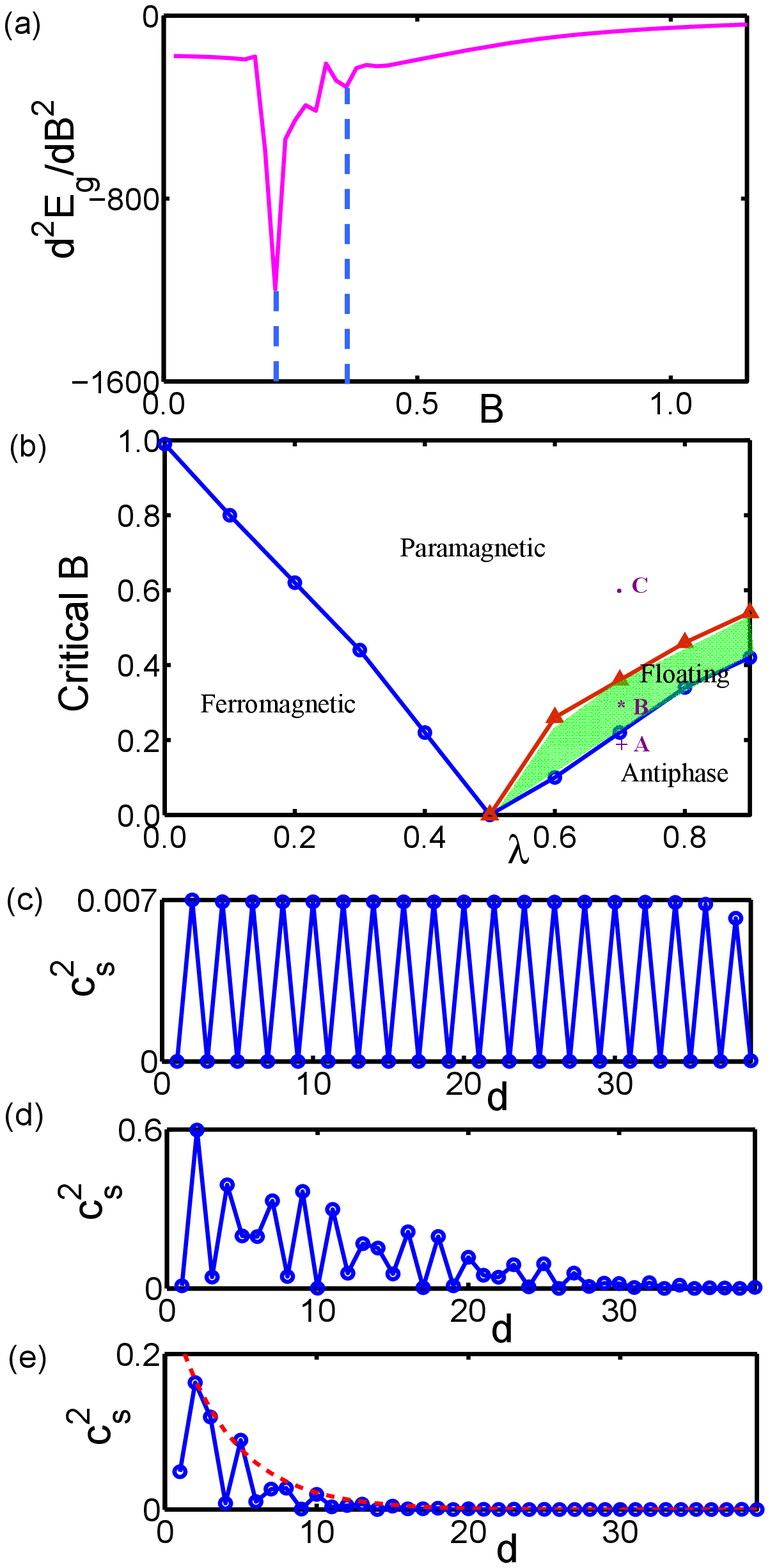}}
\caption{(Color online) (a) Fine structures and small dips in the
$dE_g^2/dB^2$ curve with the parameters $\lambda=0.7$ and $N=60$.
(b) The phase diagram of the ANNNI model. (c) The spin-spin
correlation function $c^2_s(d)$ versus the spin separation $d$ at
point A ($\lambda=0.7$ and $B=0.2$) in the phase diagram. (d)
$c^2_s(d)$ at point B ($\lambda=0.7$ and $B=0.3$) in the phase
diagram. (e) $c^2_s(d)$ at point C ($\lambda=0.7$ and $B=0.6$) in
the phase diagram. \label{fp&phase_a&phase_b&phase_c&phasediag}}
\end{figure}

Using the TEBD algorithm, we calculate the ground state energy of
ANNNI model in the external field $B$. How the ground state energy
changes with the parameters in the Hamiltonian is of great
significance because it gives us important clues on the quantum
phase transition points. Considering this, we plot the second
derivative of the ground state energy $d^2E/dB^2$ in Fig.
\ref{ProPT}, for different coupling strength $\lambda$ and system
size $N$. Notice that there are dips in the curves in Fig.
\ref{ProPT}. Their positions nearly do not change with the system
size $N$ when $N$ is large enough. These dips are where quantum
phase transitions occur, and we can read from their positions the
corresponding critical field strengths at the phase transition
points. These critical parameter values allow us to construct the
system's phase diagram which is shown in Fig. 5(b). The phase
diagram is consistent with earlier results
\cite{ref:FrusModPhaDiaAgr3} obtained using different numerical
methods.

Interestingly, as shown in Fig. 5(a) we find that in the strongly
frustrated regime there can be a segment in the curve of
$d^2E_g/dB^2$ where there are multiple extra small dips in addition
to the main dip. This region is labeled by the green area in the
phase diagram in Fig. 5(b). It is roughly at the location of the
floating phase obtained in earlier work
\cite{ref:FrusModPhaDiaAgr3}. To further clarify the characteristics
of the system in this region, we calculate the spin-spin correlation
function
\begin{equation}
c_s(d)\!=\!\left\langle \!\sigma _{N/2+1}^z\sigma
_{N/2+1+d}^z\!\right\rangle \!-\!\left\langle\! \sigma
_{N/2+1}^z\right\rangle\! \left\langle\! \sigma
_{N/2+1+d}^z\!\right\rangle.
\end{equation}
%as a function of spin separation
in this region (point B in Fig. 5(b)) and compare it to the results
in the antiphase and paramagnetic phase (point A and C in Fig.
5(b)). The results are plotted in Figs. 5(c), 5(d) and 5(e). As can
be seen in the plots, the spin-spin correlation function at point
$A$ (Fig. 5(c)) exhibits perfect long-range order which is
characteristic of the antiphase. At point C in the paramagnetic
phase, the spin-spin correlation function (Fig. 5(e)) decays
exponentially with spin separation. At point B in the green area in
the phase diagram, the spin-spin correlation function (Fig. 5(d))
appears to decay algebraically which is indicative of the floating
phase.

\section{simulating the ANNNI model in the six-junction array system}
The achievable scale of an experimental simulation system is
limited, by both decoherence and the requirement for the two-level
approximation to be valid (see appendix A). In order to study the
feasibility of our Josephson circuit system for simulation of
frustrated physics, we examine a small system to see how close its
ground state is to certain phases in Fig. 5(b). This information
will help us determine if it is possible to study the essential
characteristics of a frustrated spin system using a quantum
simulator of limited size.

\begin{figure}[t]
{\includegraphics[width=0.8\columnwidth]{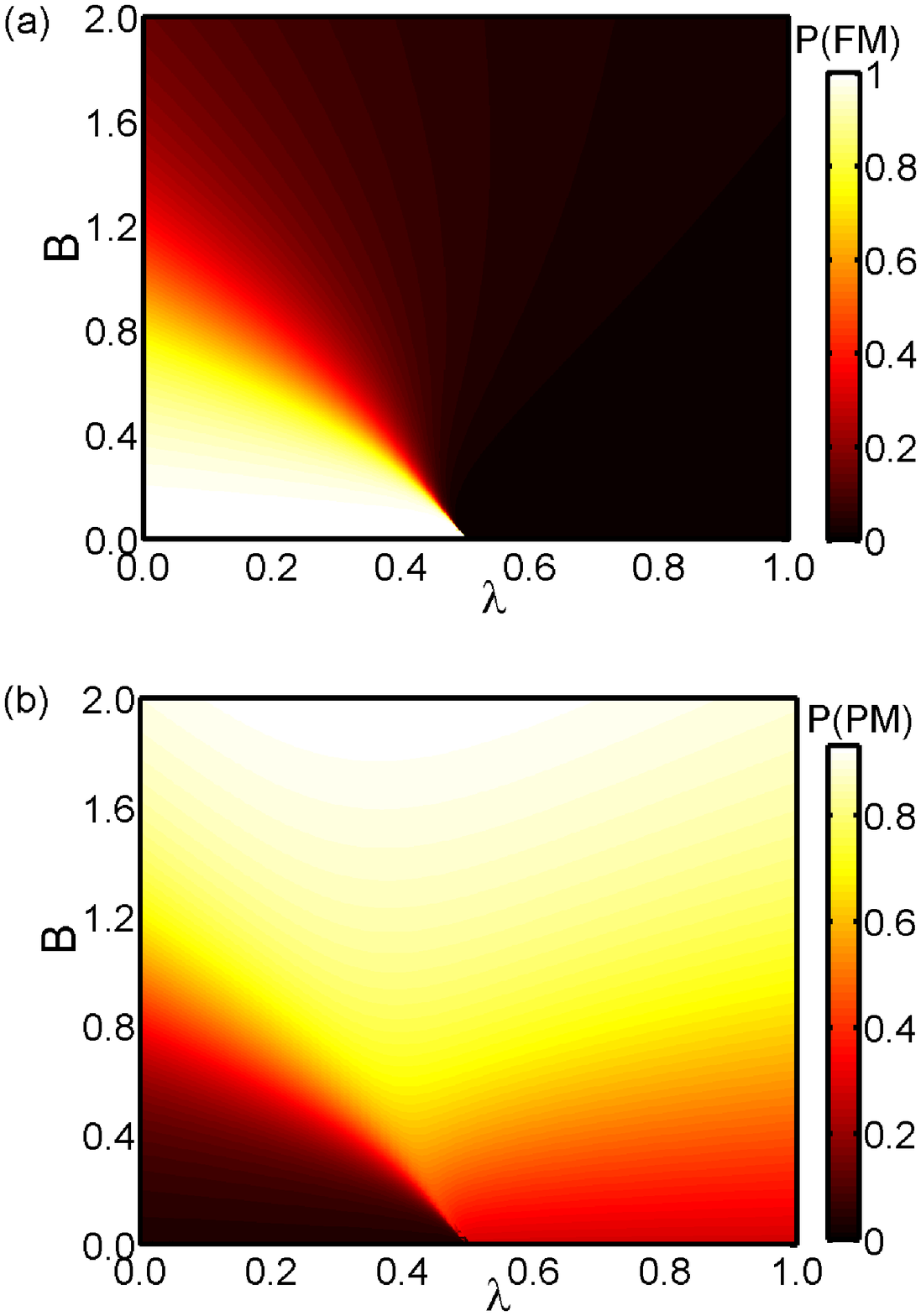}}
\caption{(Color online) (a) The probability of the ground state
$|\psi\rangle_g$ of the 6-site Josephson junction array being the
ferromagnetic state. (b) The probability of $|\psi\rangle_g$ being
the paramagnetic state. \label{site6Occu}}
\end{figure}

Taking a 6-site Josephson junction array as an example, we obtain
the ground state $|\psi\rangle_g$ of the corresponding $N=6$ ANNNI
model using exact diagonalization. Such a short chain is
insufficient to exhibit the characteristics of the floating phase,
therefore we will focus on the ferromagnetic, paramagnetic and
antiphase phases. We calculate the probabilities of $|\psi\rangle_g$
being the ferromagnetic and paramagnetic states,
\begin{eqnarray}
P(FM)&=&|_g\langle\psi|\downarrow\downarrow\cdots\downarrow\downarrow\rangle_z|^2\!\!+\!\!|_g\langle\psi|\downarrow\downarrow\cdots\downarrow\downarrow\rangle_z|^2,\label{FMOccu}\\
P(PM)&=&|_g\langle\psi|\uparrow\uparrow\cdots\uparrow\uparrow\rangle_x|^2.\label{PMOccu}
\end{eqnarray}
The results are plotted in Fig. \ref{site6Occu} for different values of $B$ and
$\lambda$.

We can see in Fig. \ref{site6Occu} that there exists a clear
junction point $\lambda=0.5$, which agrees well with the result in
Fig. 5(b). If we associate the high probability regime with the
corresponding phase, we can see that the phase regimes are nearly
the same as well. This indicates that the 6-site example can already
reveal some essential properties of the ANNNI model.
\begin{figure}[t]
{\includegraphics[width=0.8\columnwidth]{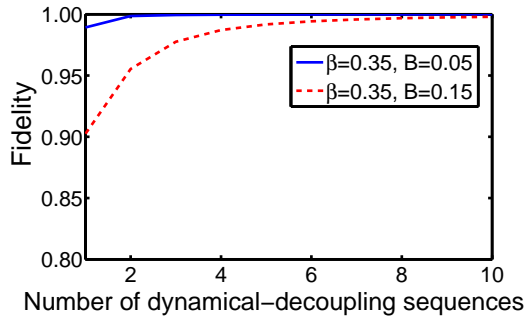}} \caption{
The fidelity of the system state to the state of the corresponding
ANNNI model, for different number of dynamical decoupling control
sequences. \label{ddctrleff}}
\end{figure}

The results in Fig. \ref{site6Occu} are based on exact
diagonalization of the $N=6$ ANNNI Hamiltonian. In order to obtain
the ANNNI model from the Hamiltonian of the Josephson junction
array, dynamical decoupling pulse sequences need to be applied to
the qubits as shown in Fig. \ref{dd}. To study the error of the
pulse engineered ANNNI model, we evaluate the evolution of the
Josephson junction array system in a time of $T=\pi$ under the
control pulses, and compare it to that of a strict ANNNI model in
the same amount of time. The initial state of the system is set to
the maximal superposition state
$|\Psi_0\rangle=\bigotimes\limits_i(|\uparrow\rangle_i+|\downarrow\rangle_i)
/\sqrt{2}$. We divide the total time $T$ into $m$ identical time
interval: $T=m \delta t$. In each interval $\delta t$, the dynamical
decoupling sequences shown in Fig. \ref{dd} are applied. In Fig.
\ref{ddctrleff}, the fidelities of the Josephson junction array
system state compared to the state of the corresponding ANNNI model
is plotted. We see that the fidelities increase with the number $m$
of decoupling sequences, and they already reach $95\%$ within a few
sequences.

We can further evaluate the effectiveness of our pulse control
scheme by comparing how the ground state changes with the external
field $B$ in a strict 6-site ANNNI model and in a pulse sequence
controlled 6-site Josephson junction array. We simulate this process
by adiabatically changing $B$ in time and calculating the
probability of the system ground state being in the ferromagnetic
phase. Here, we set $\lambda=0.4$. In Fig. 8(a), we calculate the
state evolution of a 6-site ANNNI system initially in the
ferromagnetic state $|\downarrow\downarrow\cdots\rangle_z$ when
$B=0$. We gradually increase the magnetic field $B$ from 0 with
velocity $v=0.002$, and plot the probability that the system remains
in the original ferromagnetic state at different values of $B$. We
can see that when $B$ becomes large, the system has deviated from
the ferromagnetic state, indicating that it has changed to a
different phase. In Fig. 8(b), we carry out the same study on the
6-site Josepshon junction array under the control pulse scheme in
Fig. \ref{dd}. As can be seen, the probability of the system remains
in the ferromagnetic state is almost identical to that in the
corresponding ANNNI model when the number of control sequences is
sufficient. These studies show that our control pulse based scheme
can be used to accurately simulate the frustrated ANNNI model.

\begin{figure}[t]
{\includegraphics[width=0.8\columnwidth]{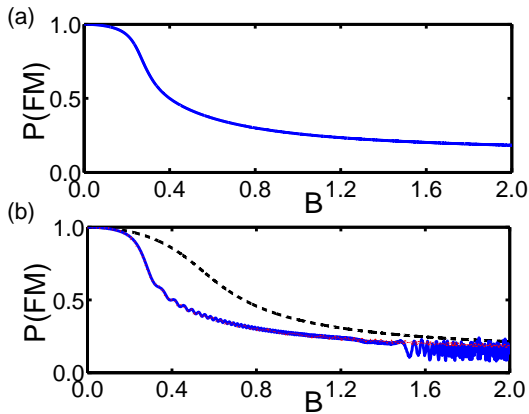}}
\caption{(Color online) (a) Phase Transition process of the ANNNI
model with $\lambda=0.4$. (b) Phase transition process of the
Josephson junction array system with and without the
dynamical-decoupling control sequence. The black dashed curve is the
result without dynamical-decoupling control sequences. The solid
blue and dotted red curves are the results with 1 and 4 control
sequences in unit time. \label{PT&DDPT}}
\end{figure}

\section{conclusion}
In conclusion, we have shown how to simulate frustrated spin
models using strongly coupled Josephson junction array. We find
that the system Hamiltonian is exactly solvable beyond the small
coupling limit, and we design a dynamical decoupling scheme to
engineer the Hamiltonian for quantum simulation of the ANNNI
model. We calculate the phase diagram of the system numerically
using the TEBD method, and demonstrate that our control pulse
based scheme can be used to simulate the corresponding ANNNI model
accurately.

\section{Acknowledgment}
This work was funded by National Natural Science Foundation of
China (Grant Nos. 11174270, 60836001, 60921091), National Basic
Research Program of China (Grant Nos. 2011CB921204, 2011CBA00200),
the Fundamental Research Funds for the Central Universities (Grant
No. WK2470000006), and Research Fund for the Doctoral Program of
Higher Education of China (Grant No. 20103402110024). L. -H. Du
and Z. -W. Zhou thank Man-Hong Yung for fruitful discussion. Z.
-W. Zhou gratefully acknowledges the support of the K. C. Wong
Education Foundation, Hong Kong.

\appendix
\section{Conditions for two-level approximation}
In deriving the spin system Hamiltonian in Eq. (\ref{eq:Spin-H}), we
used the two-level approximation for the Josephson qubits which kept
only the $n_i=0,1$ states. We study the conditions for the two-level
approximation to remain valid in this Appendix.

The Hamiltonian for the Josephson junction array system considering
contributions from all charge states reads
\begin{eqnarray}
H_n&=&\frac{1}{\lambda}\sum\limits_{i=1}^{N}(n_i-1/2)^2\nonumber\\
   & &+\sum_i\sum\limits_{j=1}^2(\lambda)^{j-1}(n_i-1/2)(n_{i+j}-1/2)\nonumber\\
   & &-B\sum\limits_i\sum\limits_n(|n\rangle_i\langle n+1|+h.c.),
\label{eq:A1}
\end{eqnarray}
where terms in the first line are the on-site charging energies of the
Josephson qubits, terms in the second line are coupling energies between
qubits, and terms in the third line are the Josephson tunneling energies. When
the effective magnetic field $B$ is nonzero, the Josephson tunneling energies
can potentially cause leakage out of the $n_i=0,1$ states and invalidate the
two-level approximation.

We take the ferromagnetic phase of the ANNNI model as an example
to estimate the probability for the qubits in the system to escape
the $n_i=0,1$ states. Initially, assume $B=0$ and the system is in
the ferromagnetic state $|\Psi\rangle_g=|0101\ldots\rangle$
(recall that there has been a canonical transformation applied on
the even sites.). When $B$ increases , the qubits can make
transitions out of the $n_i=0,1$ states. We examine the system
states that result when one of the qubits originally in the $n=1$
state changes to the $n=2$ state, because they are closest to
$|\Psi\rangle_g$ in energy among all states that violate the
two-level approximation. We denote such states $|\Psi\rangle_n$.
Their energies differ from that of $|\Psi\rangle_g$ by $\Delta
E=2(1/\lambda-1+\lambda)$ according to Eq. (\ref{eq:A1}).

According to the first order perturbation theory, the ground state
with a nonzero magnetic filed is
\begin{eqnarray}
|\Psi'\rangle_g&=&|\Psi\rangle_g+\sum_n\!'\frac{
_n\langle\Psi|H'|\Psi\rangle_g}{E_g-E_n}|\Psi\rangle_n\nonumber\\
&=&|\Psi\rangle_g+\frac{B}{\Delta
E}(\Sigma'_n|\Psi\rangle_n).\label{newgrdst}
\end{eqnarray}

With the form of $|\Psi'\rangle_g$, the total
escaping probability out of the $n_i=0,1$ Hilbert subspace
can be estimated to be
\begin{equation}
P_{esc}= (\frac{B}{\Delta E})^2\frac{N}{2},\label{escprob}
\end{equation}
where the factor of $N$ is due to the translational symmetry.

From the Eq. (\ref{escprob}), we find that the escaping probability
is proportional to $N$. Therefore the allowable system size $N$ is
limited by the tolerable escaping probability and the amplitude of
the magnetic field. For example, if an escaping probability of $5\%$
is acceptable, and $B=0.2$, the allowable size of the Josephson
junction array is about $N=10$.

\end{document}